\documentclass[%
 reprint,
superscriptaddress,
 amsmath,amssymb,
 aps,
showkeys,
]{revtex4-2}

\usepackage{blindtext}
\usepackage{amsmath, amssymb, physics}
\usepackage{bbold}
\usepackage{verbatim}
\usepackage{booktabs}
\usepackage{graphicx}
\usepackage{dcolumn}
\usepackage{cancel}
\usepackage{bm}
\usepackage[colorlinks=true, allcolors=blue]{hyperref}

\newcommand{\hy}{\hat{y}}

\newcommand{\vbx}{\vec{\mathbf{x}}}

\newcommand{\vbbeta}{\vec{\boldsymbol{\beta}}}

\newcommand{\hbtheta}{\hat{\boldsymbol{\theta}}}

\DeclareMathOperator{\sgn}{sgn}
\DeclareMathOperator*{\argmin}{argmin}

\newcommand{\E}[1][]{\mathrm{E}_{#1}}

\begin{document}

\preprint{APS/123-QED}

\title{Analog Physical Systems Can Exhibit Double Descent}

\author{Sam Dillavou}
    \email[Correspondence email address: ]{dillavou@sas.upenn.edu}
    \affiliation{Department of Physics and Astronomy, University of Pennsylvania, Philadelphia, PA, USA}

\author{Jason W. Rocks}
    \affiliation{Dayhoff Labs, Inc. Cambridge, MA, USA }

\author{Jacob F. Wycoff}
    \affiliation{Department of Physics and Astronomy, University of Pennsylvania, Philadelphia, PA, USA}

\author{Andrea J. Liu}
    \affiliation{Department of Physics and Astronomy, University of Pennsylvania, Philadelphia, PA, USA}
    \affiliation{ Santa Fe Institute, 
Santa Fe, NM, USA}

\author{Douglas J. Durian}
    \affiliation{Department of Physics and Astronomy, University of Pennsylvania, Philadelphia, PA, USA}

\date{\today}
\begin{abstract}
An important component of the success of large AI models is double descent, in which networks avoid overfitting as they grow relative to the amount of training data, instead improving their performance on unseen data.  Here we demonstrate double descent in a decentralized analog network of self-adjusting resistive elements. This system trains itself and performs tasks without a digital processor, offering potential gains in energy efficiency and speed -- but must endure component non-idealities. We find that standard training fails to yield double descent, but a modified protocol that accommodates this inherent imperfection succeeds. Our findings show that analog physical systems, if appropriately trained, can exhibit behaviors underlying the success of digital AI. Further, they suggest that biological systems might similarly benefit from over-parameterization.
\end{abstract}


\maketitle
\section*{INTRODUCTION}

Digital neural networks (DNNs) are trained at scale using centralized computation to perform digitally-precise gradient methods. This approach has been enormously successful, but is energy-inefficient compared to biological systems, where memory, compute, and training are decentralized and co-located, and computations leverage natural primitives~\cite{mead_neuromorphic_1990} rather than digital logic. Complex and expressive analog systems with these features offer tantalizing potential advantages in energy efficiency, speed, and even fault tolerance. However, physical imperfections and device variability hamper their scaling and ability to compete with digital networks.

\begin{figure*}[t]
\centering
\includegraphics[width=2\columnwidth]{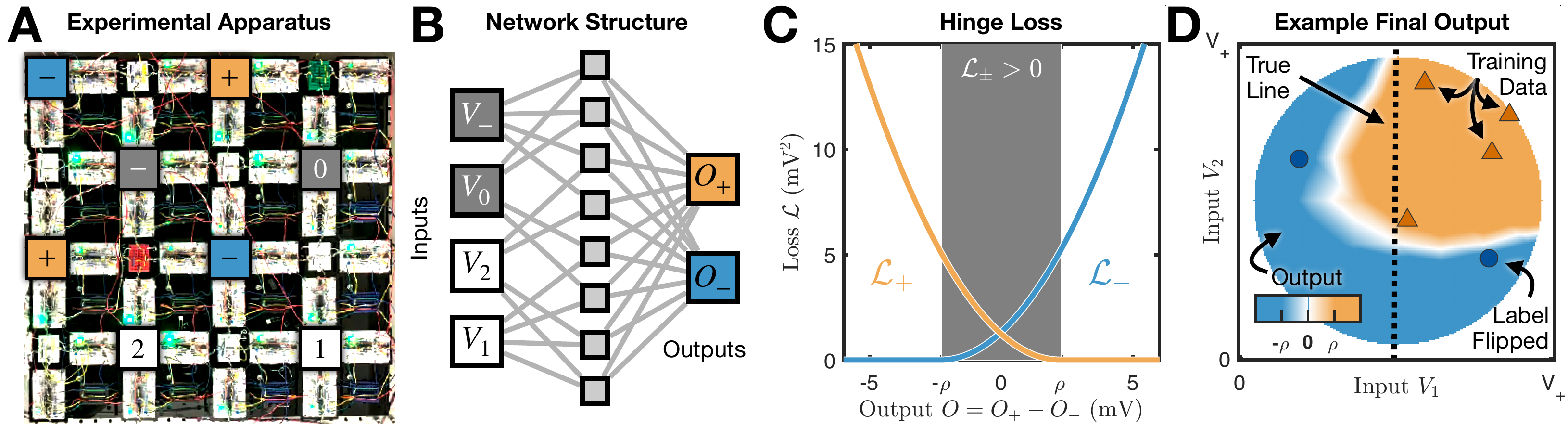}
\caption{\textbf{Network and Task Details} \textbf{(a) Experiment} Image of the Contrastive Local Learning Network (CLLN) with decorated input (gray and white) and output (blue and orange) node locations. The blue nodes labeled~$+$ are directly connected, as are the two orange nodes labeled~$-$.
\textbf{(b) Network Structure} A schematic of the self-adjusting edges (gray lines) connecting the variable inputs ($V_i$, white), constant inputs ($V_-\approx0.02$~V and $V_0\approx 0.23$~V, dark gray), and outputs ($O_\pm$, orange and blue). The smaller gray squares denote hidden nodes (unlabled in (a)). Nodes in (a) are labeled by the subscripts used in (b).
\textbf{(c) Hinge Loss Function} The losses for each class are shown. Both classes have nonzero loss within the shaded region.
\textbf{(d) Example Task Result} Results for a task with 6 training data points (circles) trained with the hinge loss in (c). The true classification division is drawn as a black dashed line. Note one label has been flipped, representing noise. Background color denotes output after training as a function of inputs $O(V_1,V_2)$. Inputs outside of the colored circle region were not used in training or testing.
}
\label{fig:1}
\end{figure*}

The extent to which analog physical systems can behave like digital neural networks is a deep question. Implications range far beyond potential applications of analog learning, including our fundamental understanding of learning, adaptability, and robustness of complex biological functionality, and the development of these capabilities in artificial physical systems. Perhaps the most counterintuitive phenomenon in modern machine learning is \textit{double descent}~\cite{rocks_memorizing_2022, belkin_reconciling_2019}, wherein generalization (test) error after training rises then counterintuitively falls as the number of individually tunable parameters (weights) approaches then surpasses the number of constraints (training data points). One might expect digital neural networks to overfit noise in the data more and more as the ratio of fit parameters to training data increases. However, in a reversal of classic statistical wisdom, once the ratio exceeds a threshold one enters a regime where generalization--the ability to treat previously-unseen (test) data--continually \textit{improves}. Thus, larger and larger systems are capable of learning more and more complex tasks. Demonstrating double descent is nontrivial in digital neural networks~\cite{rocks_memorizing_2022, belkin_reconciling_2019, schaeffer_double_2023}. Finding it in analog systems would be even more striking, and provide a bellwether of the long-term feasibility and scalability of analog physical learning networks.

To date, most analog systems that perform AI tasks use digital methods that make scaling difficult, hamstring their potential advantages, or both. Many schemes involve differentiating a digital model of the analog system~\cite{wright_deep_2022,onodera_scaling_2024}, or building an analog system to mimic digital architectures~\cite{ambrogio_analogai_2023,leroux_analog_2025,chen_demonstration_2025}, making increasingly large and expressive systems increasingly subject to what is called the ``simulation-reality gap": where device variability and imperfection lead to increasing deviation between a model and an actual physical system with increasing system size. Engineered local tuning rules for physical learning can avoid this tradeoff~\cite{stern_physical_2024}, but physical implementations typically have required digital~\cite{li_training_2024, laydevant_training_2024,xue_fully_2024,momeni_training_2025, momeni_backpropagationfree_2023} or human intervention~\cite{pashine_local_2021, altman_experimental_2024}, limiting their scalability. Notable exceptions are decentralized Contrastive Local Learning Networks (CLLNs)~\cite{dillavou_demonstration_2022,dillavou_machine_2024}. These networks of self-adjusting nonlinear resistors~\cite{dillavou_machine_2024} learn as an emergent property of their local dynamics, rather than from top-down gradient calculations. As a result, no digital model or computation is needed. Instead, the physical signals that enact the `forward' computation also encode information about the gradient, enabling individual elements to update themselves. Recent training techniques have been developed to embrace, rather than correct for, their inherent sensitivity to inevitable physical imperfections~\cite{dillavou_understanding_2025}, allowing for highly precise training.

Here we show that a nonlinear Contrastive Local Learning Network exhibits double descent as a physical phenomenon when trained with accommodation for imperfection. We encode a binary classification task with noisy data into the boundary conditions of this self-adjusting analog electronic network, and allow it to evolve (train). Its response reproduces several characteristic features of double descent as seen in digital neural networks, including a peak in response variance at the transition corresponding to a vanishing of training error. We find that digital networks with a modestly smaller number of parameters replicate the behaviors of our physical system.

\section*{Contrastive Local Learning Networks}

We use the Contrastive Local Learning Network (CLLN) shown in Fig.~\ref{fig:1}A, consisting of 32 identically-constructed edges connected via 16 nodes. In this network, physical processes naturally generate output values (`inference'), and each self-adjusting nonlinear resistive element independently enacts a local rule to tune its conductance (`training'), without centralized computation or external memory. The hardware and training procedures used in this work are described succinctly below, and in detail in~ Refs.~\cite{dillavou_machine_2024, dillavou_understanding_2025}. The theoretical framework underlying this system, Coupled Learning~\cite{stern_supervised_2021}, is closely related to Equilibrium Propagation~\cite{scellier_equilibrium_2017} and Contrastive Hebbian Learning~\cite{movellan_contrastive_1991}, but has been modified to reduce its sensitivity to inherent analog imperfection~\cite{dillavou_understanding_2025} as described below. We merge (wire together) two pairs of nodes, each on opposite sides of the periodic network, to create a connectivity pattern reminiscent of a two-layer perceptron (Fig.~\ref{fig:1}B), though with incomplete connections in the first layer. The two newly joined node pairs are used as our output nodes $O_+$ and $O_-$, and four nodes are selected as our voltage inputs, two of which, $V_-\approx0.02$~V and $V_0\approx 0.23$~V, remain always fixed, and two, $V_1$ and $V_2$, will encode our data.

\subsection*{Inference}
Consider a network of symmetric N-channel enhancement MOSFET transistors whose gate voltages $V_G$ are each determined by a charged capacitor. Provided each $V_G$ is sufficiently high, the transistors are in the triode region and obey the conductance ($G$) equation
\begin{equation}
    G = S(V_G-V_T- \overline V )
    \label{eq:conductance}
\end{equation}
where $S \approx 0.8 \times (\text{V k} \Omega)^{-1}$, $V_T \approx 0.7$~V, and $\overline V$ represents the average of the two adjacent node voltages. This dependence of conductance on voltage gives the system its nonlinearity.

When inputs (constant voltage boundary conditions) are applied to selected nodes of this network, voltages of all other nodes find a steady state (DC) value such that total dissipated power, 
\begin{equation}
    \mathcal{P} = \sum^{\text{edges}}_j G_j (\Delta V_j)^2 
\end{equation}
is minimized, with small deviations discussed in~\footnote{Due to the systems' nonlinearity, it deviates slightly from this power minimization. In practice, this has little effect on its operation, and we ignore it here. Relevant details are discussed in the supplementary information of~\cite{dillavou_machine_2024}.}. Here $\Delta V$ refers to the voltage drop across an edge (\textit{i.e.}~the source-drain voltage of each transistor). As a result, the physics in the network `calculates' a nonlinear function defined by its structure, input/output node choices, and the gate voltages:
\begin{equation}
    \mathcal{F}_{\vec V_G}(\vec V_I) = O \equiv O_+-O_-
\end{equation}
where $\vec V_I$ are the input voltages, $O$ is the output value, and $O_+$ and $O_-$ are the voltage values of two chosen nodes. 

\subsection*{Training}

The gate voltages (`weights') evolve during training according to a contrastive local rule, comparing two electronic states of the same system. In the \textit{free} state inputs are imposed, resulting in output $O$. In the other (\textit{clamped}) state, a boundary condition $O^C$ that encodes the desired output (label $L$) is enforced at the output,
\begin{equation}
    O^C = O + L_0 \ \text{sign} (L-O)
    \label{eq:clamping}
\end{equation}
where $L_0$ is a constant value. For details about implementing this clamping condition, which we call ``overclamping"~\cite{dillavou_understanding_2025}), and regarding differences among variants of the local rule (Coupled Learning, Equilibrium Propagation, and Contrastive Hebbian Learning) see Appendix~\ref{appendix:overclamping}. In practice, the two states (free and clamped) are imposed on twin networks~\cite{dillavou_demonstration_2022}, wherein gate voltages of commensurate edges are sourced from the same capacitor~\cite{dillavou_machine_2024}. These joint gate voltages are then updated via local analog circuitry~\cite{dillavou_machine_2024} enacting gradient descent on the \textit{contrastive function}, defined as the difference in power between the two states:
\begin{equation}
    \dot V_G \propto -\frac{d}{dV_G}\underset{\text{contrastive fn}}{\left[ \mathcal{P}^C - \mathcal{P}^F \right ] } = S\left[ (\Delta V^F)^2 -(\Delta V^C)^2 \right ]
    \label{eq:learningrule}
\end{equation}
where the superscript $F$ or $C$ denotes free and clamped states respectively, and $\Delta V$ is the voltage drop across the edge in question. Due to the natural power minimization of the system, the contrastive function is always non-negative, is zero only when $O=L$ (zero error), and the global gradient (LHS) can be written as a local rule (RHS). See Appendix~\ref{appendix:learningrule} for derivation of these relationships. 

We note that when using our overclamping condition (Eq.~\ref{eq:clamping}), the contrastive function (Eq.~\ref{eq:learningrule}) encodes only the sign of the error, not its magnitude. The magnitude is incorporated instead in the integration time that Eq.~\ref{eq:learningrule} is enacted during each training step, specifically 
\begin{equation}
    \Delta V_G = \int_0^{t_h} \dot V_G  \quad , \quad t_h = r_0 |L-O|
\end{equation}
where $r_0$ is a constant. This configuration greatly reduces the impact of device imperfection and variability~\cite{dillavou_understanding_2025}, enabling the results in this work. Succinctly, in Coupled Learning and Equilibrium Propagation, the difference in free and clamped states shrinks to zero with the error, ensuring that device imperfections will manifest at some level. With overclamping, that difference is held approximately constant, with a global integration time shrinking to zero instead. In practice, we adjust $r_0$ during training such that the total integration time per epoch remains constant. Details of our protocol, and results without using overclamping, are included in Appendix~\ref{appendix:overclamping}.

We ask our CLLN to perform a binary classification task in two dimensions. To enable this, we add a modification to prevent penalties for `very correctly classified' datpoints. Specifically, we set $t_h \rightarrow 0$ for any datapoint whose output value is beyond their label $L_\pm = \pm \rho$. That is, any datapoint in class $+$ is skipped in training when $O>\rho$, and similarly class $-$ datapoints are skipped when $O<-\rho$. This creates an effective shifted hinge loss, where the loss for each class is a half-parabola in $O$. We compute error explicitly as
\begin{equation}
   \mathcal{L}_\pm = (O\mp \rho)^2 \ \Theta(-\rho (O \mp \rho))
   \label{eq:hingeloss}
\end{equation}
where $\Theta$ is the Heaviside step function that zeros the loss when a datapoint is `very correctly classified'. This loss landscape is plotted in Fig.~\ref{fig:1}C. We report classification error (\%) as the fraction of datapoints with $\text{sign}(O) \neq \text{sign}(L)$.

Our 2D input data is generated from a uniform distribution within a circle in $V_1$ and $V_2$ space and labeled according to a straight-line boundary, as shown in Fig.~\ref{fig:1}D. At the start of each training run, $M$ training datapoints are selected, and the system is trained until the gate voltages $V_G$ reach a steady value, or the training error drops to zero. We vary $M$ between 1 to 64, and report averages of results across (up to) hundreds of trials. To model noisy data, we flip a random 7\% of training labels to be incorrect. We evenly distribute our test set in a grid across the same circular region in $V_1$ and $V_2$, and do not flip any test labels. Test data is interpolated to generate the smooth output field we report in background color (Fig.~\ref{fig:1}D) but test errors are calculated from actual datapoints.

\begin{figure}[h]
\centering
\includegraphics[width=1\columnwidth]{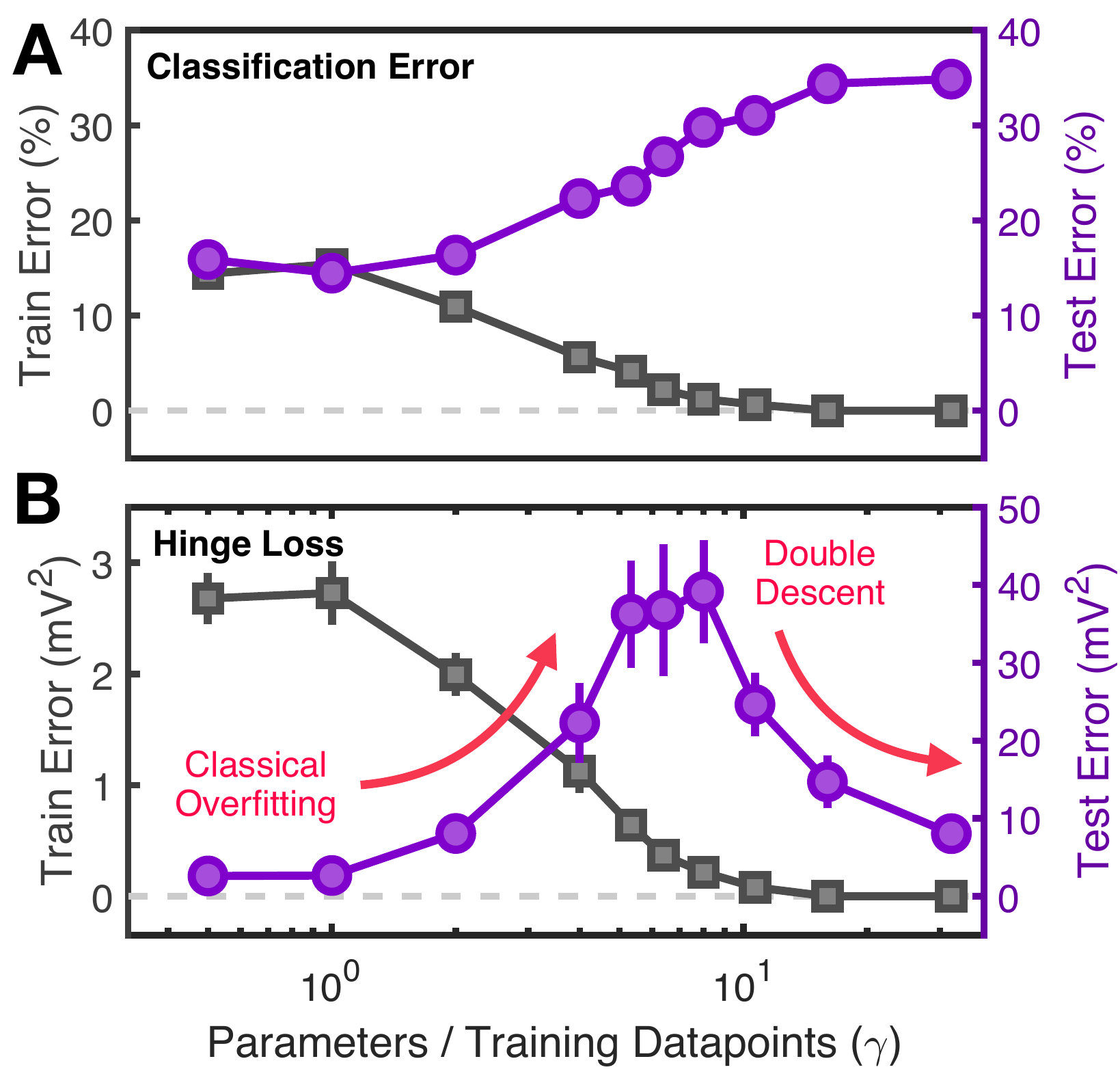}
\caption{\textbf{Experimental Results} \textbf{(a) Classification Error} at the end of training vs parameters divided by training datapoints ($\gamma =P/M$) for training (gray squares) and test (purple circles) sets. Both are monotonic functions of $\gamma$. Error bars are standard error, but are all smaller than the markers. \textbf{(b) Hinge Error} for the same experiments. Error bars are again standard error. Now, the test error is a strongly non-monotonic function of $\gamma$, with a peak at $\gamma =32/5$. }
\label{fig:2}
\end{figure}

\begin{figure*}[t]
\centering
\includegraphics[width=2\columnwidth]{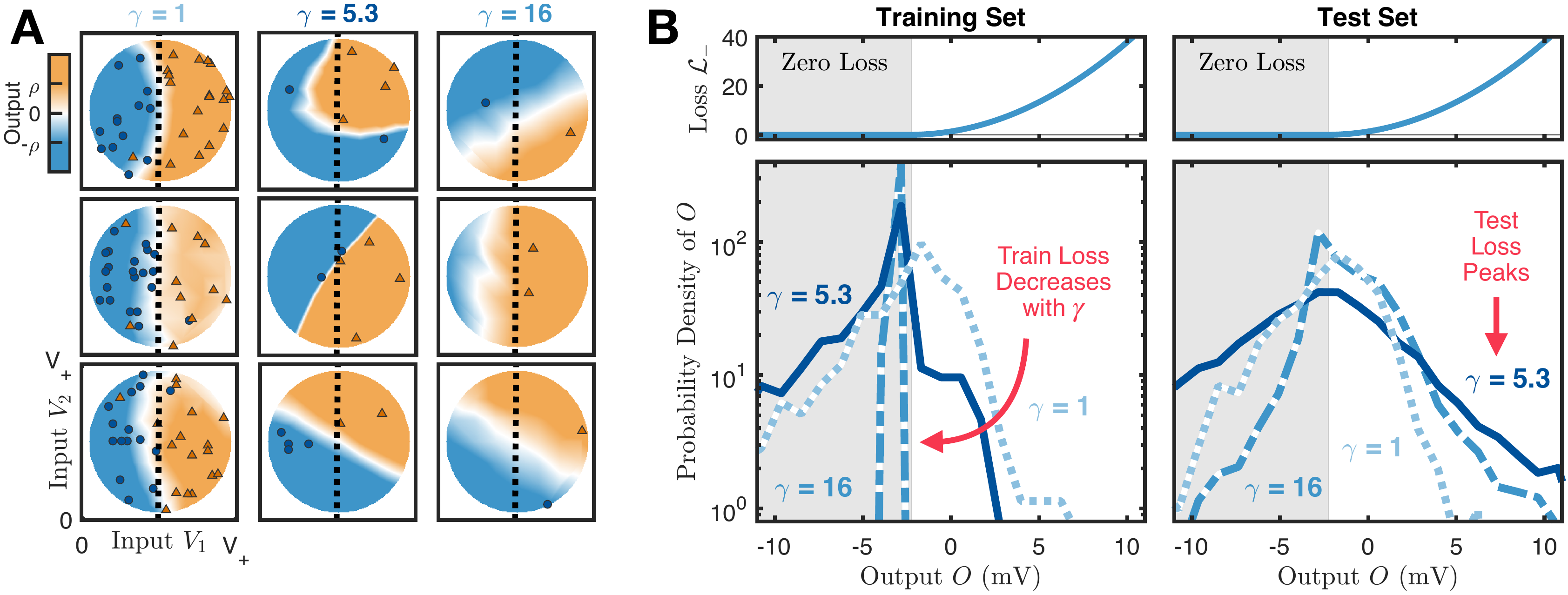}
\caption{\textbf{Output Distributions} 
\textbf{(a) Example Results} of individual training runs for $\gamma=P/M$ values of 32/32, 32/5, and 32/2. Orange triangles and blue circles are training data from the $+$ and $-$ classes respectively. The dashed line is true class division, and the background is the network output after training.
\textbf{(b) Output Distributions} Loss function (top) and final output distribution (bottom) as a function of output $O$. Data is shown only for $-$ class (blue, label $=-\rho$), equivalent plot for $+$ class is shown in Appendix~\ref{appendix:additionalresults}, Fig.~\ref{fig:S3}. Shaded regions are zero-loss for this class. Training error (left) intuitively decreases with $\gamma$ as parameters grow relative to datapoints. Test error peaks at mid-range $\gamma$, here shown by the long right-side tail for $\gamma=5.3$. }
\label{fig:3}
\end{figure*}

\section*{Experimental Results}

Our networks are fixed in size and therefore parameter count, so to study double descent we vary the number of training data points. Classification error -- the fraction of data points with the incorrect output sign ($\text{sign}(O) \neq \text{sign}(L)$) -- displays an intuitive trend. When many datapoints are used (left side of Fig.~\ref{fig:2}A), the model has enough data to find a general trend in noisy data, that is, there is enough data not to overfit. As a result the training and test errors match, signaling generalization. As fewer and fewer datapoints are used (moving right), the training dataset becomes easier to fit (training error goes down) but the system has less and less information about the underlying distribution, leading to higher test classification error (worse generalization by this metric).

However, we have constructed our training not to implement the minimization of classification error, but of a shifted hinge loss (Eq.~\ref{eq:hingeloss}, as depicted in Fig.~\ref{fig:1}C), wherein very wrong answers matter much more than slightly wrong ones. Measured in this way, the training loss looks qualitatively identical -- high when training data is plentiful, dropping to zero at small training set sizes, as shown in Fig.~\ref{fig:2}B. The test loss also intuitively rises as the number of training datapoints decreases relative to parameters (moving left to right), indicative of classical overfitting. Strikingly however, the test hinge loss displays a peak before dropping sharply as parameters greatly exceed training data points. This counterintuitive phenomenon is \textit{``double descent"}; it results from a spike in output variation at an overfitting transition. Notably, our system does not display this peak when trained without overclamping~\cite{dillavou_understanding_2025}. The fine-tuning required to fit training datasets at the transition is not feasible when the error signal (difference between clamped and free voltages) shrinks to zero with the error. Results without overclamping are included in Appendix~\ref{appendix:additionalresults}.

Individual training runs, as shown in Fig.~\ref{fig:3}A, are useful for understanding this phenomenon. First, test set classification clearly degrades with increasing $\gamma$ (decreasing training datapoints), as shown by the output (background) deviating more from the true division (dashed line). However, it is the center example (five datapoints, $\gamma = 5.3$) where the model produces the \textit{sharpest} divisions, visible by the width of the white background region. These sharp changes in output cause wrongly classified data points to contribute heavily to the hinge loss; it is as if the model is extremely confident about very wrong answers. 

This trend is also apparent when output statistics are compiled across many experiments. Viewing only the ``$-$" class (blue) for simplicity, we plot the probability density of the final output $O$ for the training set and test set in Fig.~\ref{fig:3}B. An identical figure for the ``$+$" class is included in Appendix~\ref{appendix:additionalresults}. For highly over-parameterized runs ($\gamma=16$), the training set output is sharply peaked just inside the zero-loss region; all outputs for this class are nearly identical. At this level of over-parameterization, reducing training loss to zero is simple and rapid for the network, which is rarely required to draw sharp lines, as seen in Fig.~\ref{fig:3}A (right). 

In contrast, as over-parameterization decreases ($\gamma = 5.3$), the task of perfectly fitting all 6 training data points becomes more taxing. The network draws sharper and more erratic lines, as shown in Fig.~\ref{fig:3}A (middle), in order to keep training loss low, as shown by small non-zero training loss density for $\gamma=5.3$ (Fig~\ref{fig:3}B, dark solid line). However this success does not generalize, and these sharp, varied lines cause a long-tailed test loss distribution, as shown in Fig.~\ref{fig:3}B (right side, dark solid line). 

As over-parameterization is lowered even further ($\gamma=1$), the line required to perfectly fit the training data becomes more jagged and extreme, beyond the expressivity of our system. As a result, it finds the optimal compromise in blurrier decision boundaries (less `confidence'), increasing the number of training data points with nonzero loss while reducing output variance and thus the number of `very incorrect' test data points, as shown in Fig.~\ref{fig:3}B (pale dashed lines). This represents the classic statistical tradeoff between fewer (relative) parameters and less overfitting. 

Notably, the double descent peak in our experiments occurs at $\gamma=32/5 = 6.4$. In the limit of a large network, many data points, and many input dimensions, the transition for regression occurs at $\gamma=1$, and for binary classification at $\gamma = 1/2$~\cite{gardner_space_1988}. In brief, the transition occurs when the number of constraints and parameters are equal, and in binary classification there is a $1/2$ probability that each data point (constraint) lies in the zero hinge loss region, rendering it irrelevant. Thus a plausible hypothesis for the high $\gamma$ transition value is an effectively lower number of parameters of our system; were we to have 2-3 parameters, the transition at 5 datapoints would yield $\gamma = 1/2$. However this is clearly an underestimation of the nonlinear curve-fitting abilities on display in Fig.~\ref{fig:3}. Instead, via comparison to digital networks, we now argue that finite-size effects and parameter restrictions (including non-negativity) in our system account for this unexpected transition location. 

\section*{DIGITAL COMPARISON}
To understand why the peak in the double descent curve occurs at $\gamma \approx 5.3$ instead of $\gamma=1/2$, we carry out analogous trials with very small digital neural networks. Specifically, 
we train an ensemble of digital nonlinear classifiers on a 2D binary classification dataset generated in the same manner as our experimental data, using a shifted hinge loss. The classifiers are two-layer perceptrons with either rectified linear units (ReLU) or hyperbolic tangent (Tanh) as their nonlinear activations. Network inputs mirror the physical experiments: two variable inputs and a third, constant input. This final input is added because we do not permit these networks to adjust bias values (to match our experiment). Note that the experiments required a fourth (constant) input to serve as a voltage reference. The first layer of these digital networks is randomly chosen and frozen: nonlinear features to be fit by the trainable second layer. The second layers' size is reported as parameter number. For each choice of nonlinearity, we selected a single (Ridge) regularization value to approximately match the test loss peak height. Reported values are averaged across 90,000 training runs. Further details are reported in Appendix~\ref{appendix:digitaldetails}.

\begin{figure}[h]
\centering
\includegraphics[width=1\columnwidth]{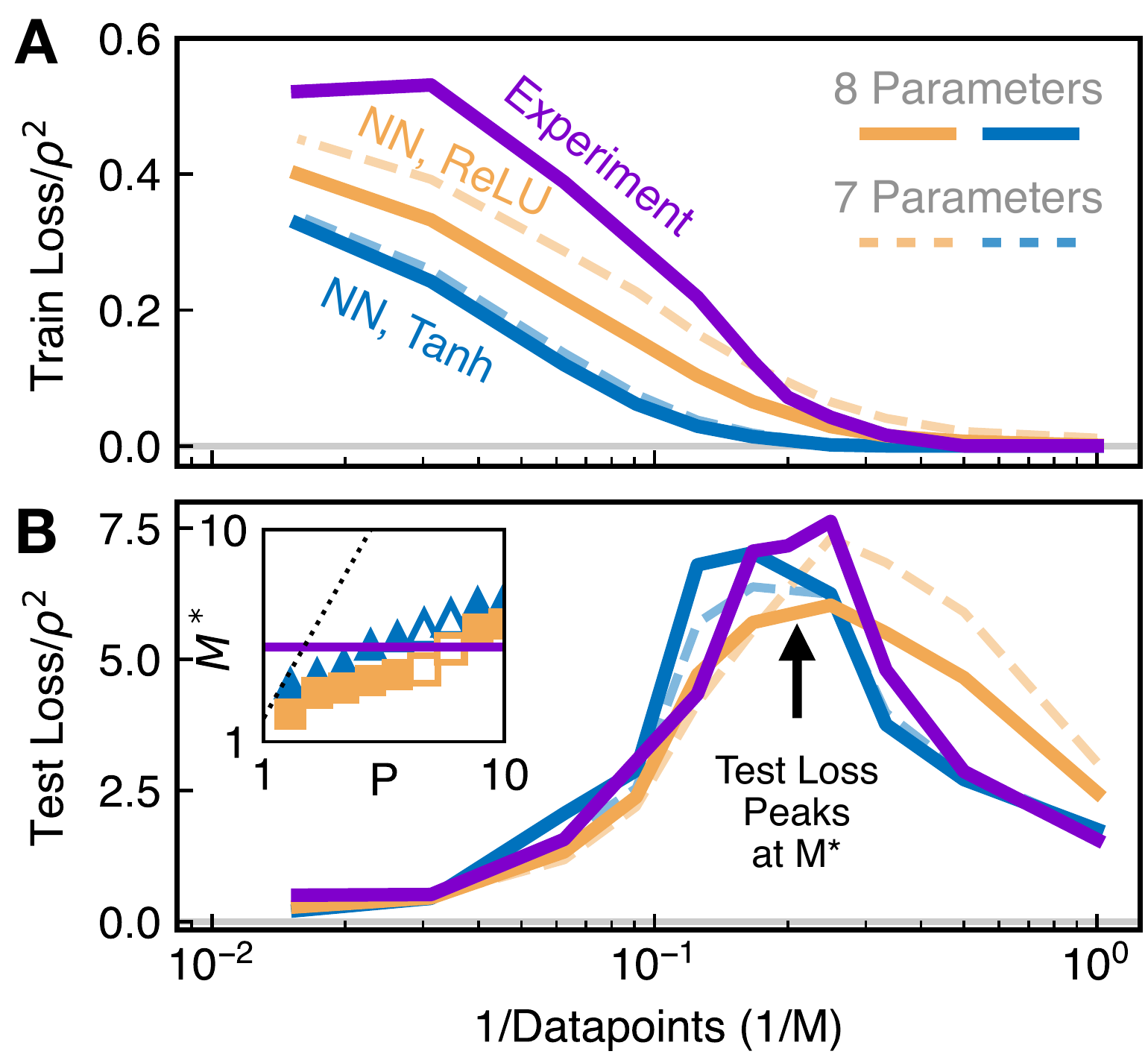}
\caption{\textbf{Comparison with Digital Networks} \textbf{(a) Training Hinge Loss} vs 1 / data points ($1/M$). Experimental results (purple) are compared with average training loss for an ensemble of neural networks with 8 (solid lines) and 7 (dashed lines) trainable parameters ($P$). Networks with ReLU (yellow) and tanh (blue) activations were used. These networks were trained on a two-feature (2D input dimension) binary classification task generated in the same manner as the experimental data.
\textbf{(b) Test Hinge Loss} vs 1 / datapoints for the same task and networks. $M^*$ denotes the test loss peak. \textbf{Inset:} $M^*$ vs parameters $P$ for ReLu (yellow squares) and tanh (blue triangles) networks. Networks with 7 and 8 parameters (curves in (a) and (b)) are highlighted. The horizontal purple line denotes the experimental $M^*$ value, and the black dotted line denotes the theoretical (large-scale) prediction $M^*=2P$
}
\label{fig:4}
\end{figure}

We find the best qualitative agreement between 8- or 7-parameter digital neural networks and our analog experiment, as shown in Fig~\ref{fig:4}A and B. Note that experimental results here are the same as those in Fig.~\ref{fig:2}B, but are plotted against 1/datapoints ($1/M$) alone, rather than $\gamma=P/M$, so that we compare curves with different $P$ values. While the experiment returned slightly higher training losses (Fig~\ref{fig:4}A), the test loss curves of the digital and experimental systems approximately match in both shape and height (Fig~\ref{fig:4}B). By fitting the test loss peak to a parabola, we extract the peak location $M^*$ for a range of parameter values (Fig~\ref{fig:4}B, inset). These differ significantly from the theoretical prediction $M^* = 2P$ or $\gamma = P/M^* = 1/2$ (dashed line).

These results indicate that the feature dimension (2D inputs) plays a significant role in the peak location of our experiments. Larger digital neural networks on higher-dimensional binary classification tasks display the double descent peaks at the predicted location $\gamma \approx 1/2$, as shown in Appendix~\ref{appendix:digitaldetails}. For the simulations in the main text (2D inputs) we find $\gamma \approx 2$. 

To reconcile the digital neural network result with our experimental results ($\gamma\approx 5$), we note that parameters in the physical network are bounded, most notably to be positive only. This is why our system's output is considered to be a difference of two nodes, effectively allowing positive parameters (conductances) to yield negative input-to-output relationships. This effectively halves the parameter count, explaining one factor of $2$ in $\gamma$. Further, for the transistors to remain in the desired operation regime, gate voltages are restricted within approximately one decade. Nonlinear effects give the system access to a conductance range likely closer to two orders of magnitude~\cite{dillavou_machine_2024}, but this is still a strong additional restriction which could easily account for the remaining discrepancy. This bounding acts as an effective regularizer, preventing extreme values known to occur during overfitting~\cite{krogh_simple_1991}.

\section*{DISCUSSION}

We have demonstrated that analog physical systems, in the absence of any digital computation, can display the phenomenon of double descent. The appearance of this fragile signature suggests that even imperfect analog hardware can be utilized to perform precise, many-parameter optimization. Further, it suggests that the passage of gradient information via physical processes is a valid and promising alternative to digital optimization of a system-wide model of hardware (a ``digital twin"), the tactic used (sometimes implicitly) in nearly all analog learning hardware~\cite{momeni_training_2025}.

Creating a digital twin requires extremely precise and thus computationally and/or temporally expensive adjustment or feedback of the model or hardware. In the experiments described in this work, it is a common occurrence to have a gain of 1 between some gate voltages and the network output ($dO/dV_G \approx 1$). This suggests that accurate training using a digital twin would require a model with generic accuracy better than 1~mV, or equivalently within $S \times 1~\text{mV} \approx 0.8 \mu\text{S}$ in conductance, of order 0.1\% accuracy. While this is certainly possible, it requires feedback or extensive measurement (as in \textit{e.g.}~\cite{gokmen_algorithm_2020}), as tolerances below 10\% on many MOSFET parameters can be difficult to achieve. Digital twins of electronic components typically also assume a simple form for each element, allowing small deviations (like nonlinearities in conductance) to compound across a network. To avoid this kind of compounding, even more extensive data-driven modeling~\cite{wright_deep_2022} or zero-order methods~\cite{mccaughan_multiplexed_2023} with less favorable scaling properties have been introduced.

Our system avoids this penalty by leveraging the actual physical signals of the network to implicitly calculate gradients. While a model of each element is assumed (Eq.~\ref{eq:conductance}), deviations from this model do not compound with one another. An incorrectly modeled adjacent MOSFET does not affect the update calculation of a given edge; that calculation depends only on the \textit{actual} electronic signals that propagate through the network. In this sense our system is closer to digital neural networks than other analog hardware, in that the model \textit{is} the system. Notably, the twin-networks in our setup are `models’ of each other in this way, but can take advantage of paired transistors with tighter intra-pair tolerances. Single-network implementations of similar frameworks would avoid even this discrepancy, and some have been proposed~\cite{pashine_local_2021,oh_memristor_2023,anisetti_frequency_2024,falk_temporal_2025}, though not yet experimentally implemented.

Because modeling discrepancies and other hardware imperfections (like measurement or update rule errors~\cite{dillavou_understanding_2025}) sandbag a system's ability to precisely tune its output, we propose using the double descent transition as a hardware benchmark. Rather than stated parameters, it is instructive to understand how many 'effective' parameters a hardware system has, that is, how does its performance compare to digital neural networks of varying size.

The `overclamping' technique~\cite{dillavou_understanding_2025} leveraged in this work suppresses analog measurement errors that each component makes during training. By keeping the error propagation signal finite (but allowing integration time to shrink with error), the required fine-tuning of each analog element can be achieved. In a sense, this technique `zooms in' on the loss landscape, amplifying error signals above measurement error. However there is no free lunch, and there is still a tradeoff between signal size (large nudge) and gradient accuracy (smaller nudge). In our experimental setup this has not caused problems, but investigating this tradeoff at scale, as well as determining how to mitigate its effects, is a fruitful subject for future work. Were it to scale successfully, integrating physically-propagated error signals through self-adjusting elements might prove useful in constructing precise compound analog systems from sensors to quantum computers.

It has been suggested that the transition from the under- to over-parameterized regime is akin to a jamming transition~\cite{spigler_jamming_2019}. Another theory~\cite{li_broken_2025} of the transition suggests that it is analogous to the normal-superconducting transition in London theory. Generalizing these ideas to finite systems and comparing to our experimental systems is an important direction for future work. 

Double descent implies a striking deviation from the normal behavior of many-body physical systems. Typically, as such systems grow in size, their properties change (``more is different") but eventually converge to a thermodynamic limit. Double descent instead implies ``many more can be more different": physical behaviors can continue to change indefinitely as the system increases in size. 

Our demonstration of double-descent in a non-digital setting has interesting implications for the general physical design of complex systems, and biology in particular, pointing to the importance of over-parameterization. We hypothesize that, as in digital neural networks, over-parameterization may underlie many complex, emergent collective behaviors in biological systems, rendering them easy to tune for function, vastly more expressive, and robust and adaptable to varying environmental conditions. Evidence of overparameterization has been observed in computational models tuned for biologically-inspired function~\cite{rocks_designing_2017,rocks_limits_2019,tah_minimal_2025,galvanicunha_building_2024,wang_mechanosensitive_2025,arzash_rigidity_2025}, suggesting that biology has taken advantage of these features. The notable gap between the physical abilities of humans and robots may be partially explained by this point; human bodies physically reconfigure as physical skills are learned. Notably, the mathematical frameworks underlying our work~\cite{stern_supervised_2021, dillavou_understanding_2025} are extendable to mechanical systems~\cite{altman_experimental_2024}. Such a system might permit new levels of physical grace in robotics, even potentially unlocking new forms of physical functionality, akin to the revolution in digital abilities enabled by neural networks.

\acknowledgements 
This work was supported by the National Science Foundation through grants MRSEC/DMR-2309043 (DJD/AJL) and DMR-MT-2005749 (AJL), as well as by Simons Foundation through Investigator grant \#327939 (AJL). SD acknowledges support from the Data Driven Discovery Initiative, the Center for Soft and Living Matter at the University of Pennsylvania, and the UK Advanced Research + Invention Agency (ARIA) under project NACB-SE02-P06.

\bibliography{bib,bib_other}

\appendix

 \section{Learning Rule} \label{appendix:learningrule}

Below we derive the local learning rule implemented in our self-learning hardware. The goal of the rule below is to allow edges of our network to self-tune in response to training data consisting of paired sets of input(s) and label(s) applied as boundary conditions. When inputs only (free) or inputs and labels (clamped) are applied, node voltages $V_i ^F$ or $V_i ^F$ are generated, respectively. Our edges evolve by varying their gate voltages $V_G$, and which are kept high to obey the ohmic response of an n-channel MOSFET transistor, with conductance $G$ given by Eq.~\ref{eq:conductance} (reproduced here):
 \begin{equation}
    G = S(V_G-V_T- \overline V )
\end{equation}
where $S \approx 0.8 \times (\text{V k} \Omega)^{-1}$, $V_T \approx 0.7$~V, and $\overline V$ represents the average of the two adjacent node voltages. This dependence of conductance on voltage gives the system its nonlinearity, which is generally a secondary effect. Linear passive electronic systems (and this system, in approximation) minimize total dissipated power $\mathcal{P}$ in steady-state. Thus, we choose a governing equation that takes advantage of this fact, performing gradient descent with respect to gate voltages $V_G$ on the \textit{Contrastive Function}, the difference in powers between the clamped and free states,
 \begin{equation}
    \dot V_G \propto -\frac{d}{dV_G}\underset{\text{contrastive fn}}{\left[ \mathcal{P}^C - \mathcal{P}^F \right ] } 
    \label{eq:contrastivegradient}
\end{equation}
We note that 
\begin{enumerate}
\item The first derivative of power w.r.t.~any unconstrained node voltage (and thus voltage drop) is zero $\frac{\partial \mathcal{P}}{\partial \Delta V_k}=0$, because power is minimized.
\item Conductance $G$ values do not depend on each other
\item Power is a local sum: $\mathcal{P} = \sum_k P_k  = \sum_k G_k \Delta V_k^2$ (here $k$ is the index over edges)
\item The derivative of a conductance w.r.t.~a gate voltage $\frac{dG_j}{dV_{G,i}}$ is zero unless $i=j$ (same edge) in which case the result is $S$, with only small higher order corrections from feedback with the DC voltage state, which we will ignore.
\end{enumerate}
This allows us to greatly simplify the gradient of total power in steady state (free or clamped) w.r.t.~our gate voltages $V_G$:
\begin{equation} \label{makelocal}
   \frac{d\mathcal{P}}{dV_{G,i}} = \sum_k \sum_j\underbrace{\frac{\partial P_k}{\partial G_j}}_{\delta_{kj}V_j^2}\cancelto{S \delta_{ij}}{\frac{dG_j}{dV_{G,i}}} + \cancelto{0}{\frac{\partial P_k}{\partial \Delta V_j}}\frac{d\Delta V_j}{dV_{G,i}} =S V_i^2
\end{equation}
hich allows us to rewrite eq.~\ref{eq:contrastivegradient} as a local rule:
\begin{equation} \label{gradeqn}
    \dot V_G \propto (\Delta V^F)^2- (\Delta V^C)^2
\end{equation}
as stated in Eq.~\ref{eq:learningrule}.

The difference between the Free and Clamped states is the clamped boundary conditions. In standard Coupled Learning, this boundary condition is defined as
\begin{equation}
    O^C = (1-\eta)O^F + \eta L
    \label{eq:standardclamping}
\end{equation}
where $O^F$, $O^C$, and $L$ are the free and clamped outputs, and labels, respectively.  Here $\eta$ is a hyperparameter (nudge factor), with $\eta=1$ corresponding to simply clamping the label $L$ directly, and $\eta<1$ `nudging' towards the label. Because power $\mathcal{P}$ is approximately minimized in each case, the clamped power must be greater than the free power, unless the clamped boundary condition is already satisfied in the Free state. This happens only when the network naturally generates the label(s) (desired values) at the output(s) ($O^F=L$), meaning the error is zero. Thus, the minima of the contrastive function $\mathcal{P}^C-\mathcal{P}^F$ are the same (zero-error) minima as a standard cost function.

\begin{figure}[t]
\centering
\includegraphics[width=1\columnwidth]{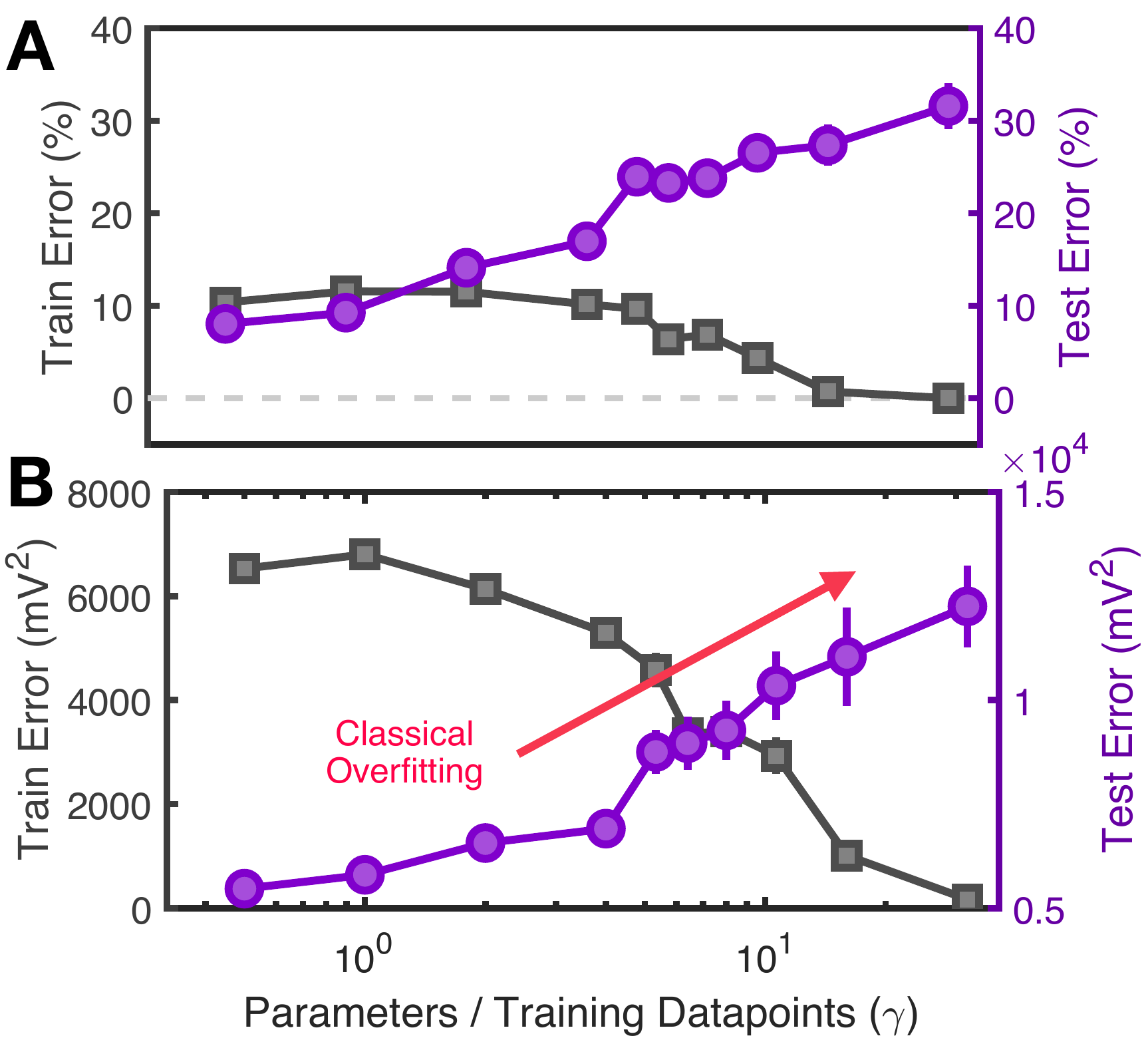}
\caption{\textbf{Typical Experimental Results Without Overclamping} \textbf{(a) Classification Error} at the end of training vs parameters divided by training datapoints ($\gamma =P/M$) for training (gray) and test (purple) sets. Error bars are standard error. \textbf{(b) Hinge Error} for the same experiments vs training datapoints for training (gray) and test (purple) sets. Error bars are standard error. Note that here $\rho \approx 11$~mV, and similarly shaped curves result from larger labels. Labels as small as those used in the main text ($\rho \approx 2.3$~mV) cannot be meaningfully distinguished through training without overclamping.}
\label{fig:sup1}
\end{figure}

 \section{Overclamping} \label{appendix:overclamping}
The clamping protocol described above has been used successfully~\cite{dillavou_demonstration_2022,wycoff_desynchronous_2022,stern_physical_2022,dillavou_machine_2024} but routinely hits an accuracy ceiling. The underlying reason, explored in depth in~\cite{dillavou_understanding_2025}, is that the contrast between the Free and Clamped states (effectively the learning signal) decays to zero with the error. Thus, once error falls below a certain level, hardware and measurement imperfections dominate over the signal, hamstringing expressivity of the network. Notably, this is also a problem in many related frameworks, including Equilibrium Propagation~\cite{scellier_equilibrium_2017}, which explicitly operates in the `infinitesimal nudge' limit, where the contrast signal is vanishingly small.

A solution to this problem is \textit{overclamping}~\cite{dillavou_understanding_2025}, a technique employed in this work. Instead of varying the clamped voltage condition in proportion to the error, the distance between the free and clamped outputs is kept at constant magnitude $L_0 = |O^F-O^C|$.  The gate voltage in our circuit is modified by generating a current proportional to the learning rule~\ref{eq:learningrule} and integrating for a time $t_h$. In standard Coupled Learning, $t_h$ is constant, but here we vary this integration time in proportion to the error $t_h \propto |O^F-O^C|$, recovering the desired step size dependence. The resulting technique avoids small-signal problems that are catastrophic for analog hardware. As in \cite{dillavou_understanding_2025}, we normalize learning rate by keeping a running tally of average error. Training time is distributed to each datapoint such that the total learning (integration) time per epoch remains fixed; see \cite{dillavou_understanding_2025} for more details.

\begin{figure*}[t!]
\centering
\includegraphics[width=2\columnwidth]{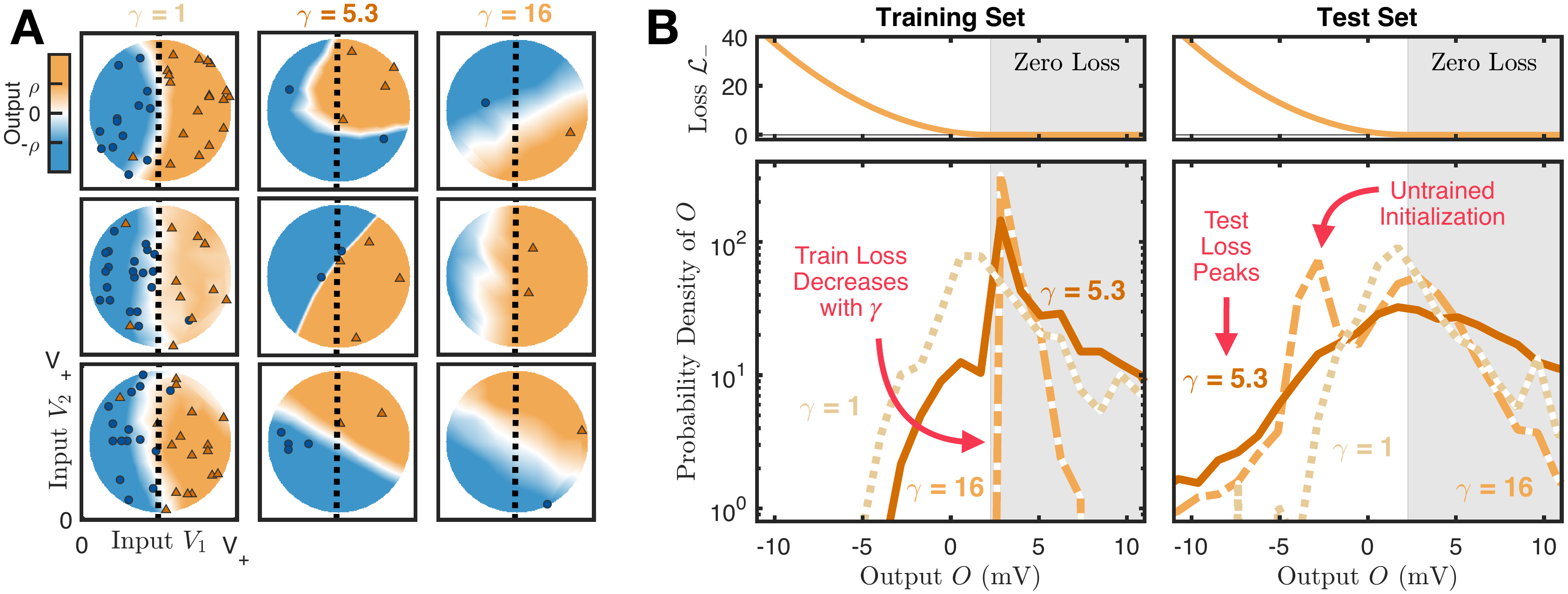}
\caption{\textbf{Output Distributions (Class $+$)} 
\textbf{(a) Example Results} of individual training runs for $\gamma=P/M$ values of 32/32, 32/5, and 32/2. Orange triangles and blue circles are training data from the $+$ and $-$ classes respectively. The dashed line is true class division, and the background is the network output after training. These are a reproduction of Fig.~\ref{fig:3}A. 
\textbf{(b) Output Distributions} Loss function (top) and final output distribution (bottom) as a function of output $O$. Data is shown only for $+$ class (orange, label $=+\rho$), equivalent plot for $-$ class is shown in main text (Fig.~\ref{fig:3}). Shaded regions are zero-loss for this class. Training error (left) intuitively decreases with $\gamma$ as parameters grow relative to datapoints. Test error peaks at mid-range $\gamma$, here shown by the long right-side tail for $\gamma=5.3$. Note that when initialized, our network outputs a slightly negative value regardless of input, creating a peak for small-datapoint runs (see text). }
\label{fig:S3}
\end{figure*}

In practice, we use a nudging technique to approximate a constant shift. Namely, we enforce 
\begin{equation}
    O^C = (1-\eta)O^F + \eta L_\pm
\end{equation}
Where $\eta = 1/4$, and $|L_\pm| \approx 450 \text{mV} \gg \rho \approx 2.3 \text{mV}$, where $\rho$ is the label and thus the approximate scale of all outputs ($O^F$). We choose the sign of $L_\pm$ based on the sign of $L-O^F$ (the direction of the desired output). The result is approximately equivalent to Eq.~\ref{eq:clamping} (reproduced here):
\begin{equation}
    O^C = O^F + L_0 \ \text{sign} (L-O^F)
\end{equation}
with $L_0 \approx 112.5 \text{mV}$

Importantly, this technique is necessary to observe double descent. Despite a wide range of choices for task, hyperparameter, input/output placement, and network architecture, we observed no convincing double descent signal without overclamping. A typical example of the curves from Fig.~\ref{fig:2} without overclamping are shown in Fig.~\ref{fig:sup1}, displaying a monotonically increasing test error curve with $\gamma$, indicative of classical overfitting. This is a consequence of less-precise adaptation; we note that the training error here is over 2000$\times$ higher than in the main text. Finally, it is worth noting that performing standard clamping requires extremely fine control of an imposed voltage (as do many learning techniques). Overclamping, in contrast, requires only fine control over a single global learning time signal, which is much easier in practice.
 
\section{Additional Results} 
\label{appendix:additionalresults}
The training results for our binary classification problem were shown only for one class ($-$) in Fig.~\ref{fig:3} to avoid clutter. The results for the other ($+$) class are shown here, in Fig.~\ref{fig:S3}, presented identically. The only notable difference between these results and those in the main text is the peak in the $\gamma = 16$ distribution, which results from the initialization conditions of our network. By chance (dependent on the architecture, input/output choices, and initial gate voltages), our network produced a small negative output, relatively insensitive to the variable inputs, at the start of each experiment. As a result, any training runs with \textit{only} datapoints in the $-$ class was immediately satisfied, creating test outputs at that value as well, regardless of class. This peak does not show up in the main text figure because the reverse does not occur; training runs with $+$ data only would shift the output to the $+$ side, but just over the line, creating a gradual decay in the distribution rather than a second peak. Whenever both classes are present (nearly always the case at higher datapoint counts), this initialization effect is trained away.

\begin{figure}[h!]
\centering
\includegraphics[width=1\columnwidth]{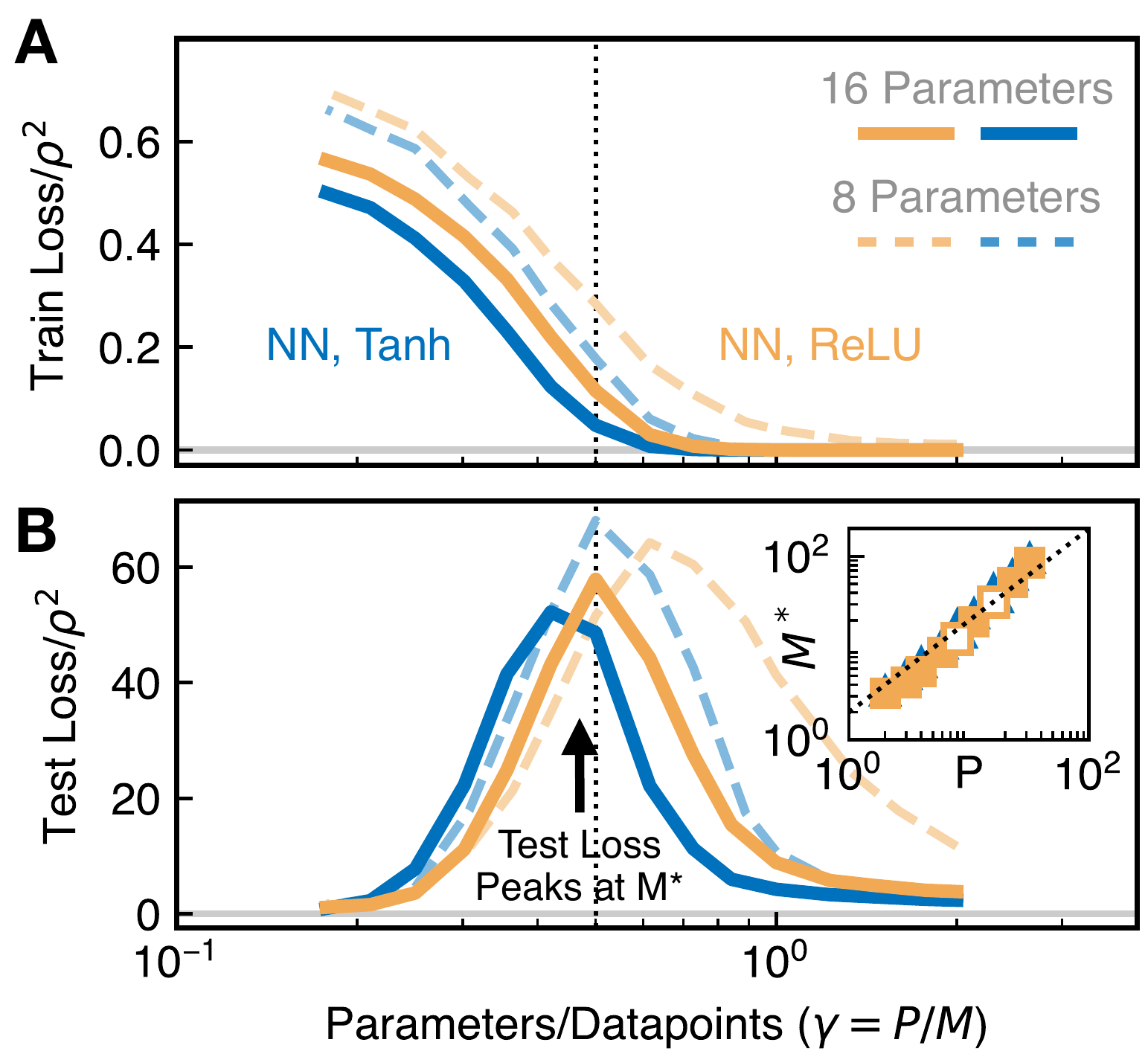}
\caption{\textbf{Results from a Higher-Dimensional Task} \textbf{(a) Training Hinge Loss} vs paramters over datapoints ($\gamma = P/M$) with ReLU (yellow) and tanh (blue) activations. Networks were trained on a 16 feature (16-dimensional) input space for a binary classification task. The vertical dashed line is at the theoretical prediction in the thermodynamic limit for the transition point ($\gamma^* = 1/2$).
\textbf{(b) Test Hinge Loss} vs $\gamma = P/M$ for the same task and networks. 
\textbf{Inset:} $M^*$ (transition location in datapoints) vs parameters $P$ for ReLU (yellow squares) and tanh (blue triangles, mostly obscured) networks of varying parameter counts. The dashed line is the theoretical prediction $\gamma^*=P/M^*=0.5$ Hollow shapes represent the curves diplayed in (a) and (b).
}
\label{fig:sup2}
\end{figure}

 \section{Training of Digital Networks}
 \label{appendix:digitaldetails}
 
As a point of comparison, we trained digital networks on a two-feature (2D) binary classification task using a shifted hinge loss. Our models were 2-layer dense networks, with the first layer's weights, matrix $W_0$, fixed. The trainable function was thus
\begin{equation}
    O(I,W_1) = \vec W_1 \cdot \sigma\left( W_0 \cdot 
    \vec I \right)
\end{equation}
where $\sigma()$ was an element-wise nonlinearity. Runs were performed both with $\sigma$ as ReLU and as tanh. To more faithfully mimic the experimental conditions, we did not allow the network to adjust `bias' parameters (which would be additive constants in the above equation), and instead added a constant input value ($I_3= 1$) as a stand in for the experiment's constant-value boundary conditions. The result is an adjustable linear mapping of nonlinear feature vectors, a system known to robustly display double descent phenomena in the thermodynamic limit~\cite{rocks_memorizing_2022}.

The two variable input values ($I_1$ and $I_2$) were drawn from a uniform 2D circular distribution (as in the experiment), here with maximum and minimum value of 1/2 and -1/2. As in the experiment, a linear class boundary is chosen, and 7\% of the labels in the training data are flipped (test data is not corrupted in this way). For each choice of parameters $P$ (length of $\vec W_1$) and training datapoints $M$ we generate 300 $W_0$ feature projections and 300 random training data sets, respectively. We then perform the 90,000 experiments corresponding to each pair, and repeat for each $P$,$M$ combination. Weights are fit using the Broyden–Fletcher–Goldfarb–Shanno (BFGS) algorithm, and (Ridge) regularization was selected to approximately match the peak heights of the test error, normalized by label magnitude squared. These values were $7\times 10^{-5}$ for tanh and  $7 \times 10^{-4}$ for ReLU.

The peak location found in the main text clearly falls well below the thermodynamic limit result of $M^* = 2P$, for both experiment and digital simulations. This is a finite-size effect in \textit{features}, which we confirm by running larger simulations with the same digital configuration, but using 16 variable inputs instead of 2. Results converge faster than with 2 inputs, so we use only 150 unique feature projections and 150 unique training data sets for each $P$ and $M$ pair, and we do not restrict ourselves to a hyper-sphere, but use a hyper-cube instead (as the relative hyper-sphere volume shrinks rapidly with dimension). The remaining protocol details are identical. The results of these runs are shown in Fig.~\ref{fig:sup2}A (training) and B (test error), and display the expected $M^* = 2P$ scaling (inset). We display these results versus $\gamma$ instead of $1/M$ to highlight the (relative) consistency of the transition location in $\gamma$ when parameter count changes, as predicted in the thermodynamic limit. The finite-size effects around this transition have not been thoroughly explored, and are a fascinating subject for future work.

\end{document}